# High-speed quantitative nanomechanical mapping by photothermal off-resonance atomic force microscopy

*Hans Gunstheimer[1,2], Gotthold Fläschner[2,†], Jonathan D. Adams[2], Hendrik Hölscher[1,3,*], Bart W. Hoogenboom[2,*]*

[1] Institute of Microstructure Technology, Karlsruhe Institute of Technology (KIT), Hermann-von-Helmholtz-Platz 1, 76344 Eggenstein-Leopoldshafen, Germany  [2] Nanosurf AG, Gräubernstrasse 12 – 14, 4410 Liestal, Switzerland

[3] Karlsruhe Nano Micro Facility (KNMFi), Karlsruhe Institute of Technology (KIT), Hermann-von-Helmholtz-Platz 1, 76344 Eggenstein-Leopoldshafen, Germany

[†] Present address: Institute for Bioengineering of Catalonia (IBEC), C/ Baldiri Reixac 10-12 · 08028 Barcelona, Spain.

[*] Corresponding authors. Email: hendrik.hoelscher@kit.edu, hoogenboom@nanosurf.com

**Abstract**

Atomic force microscopy (AFM) is widely used to measure surface topography of solid, soft, and living matter at the nanoscale. Moreover, by mapping forces as a function of distance to the surface, AFM can provide a wealth of information beyond topography, with nanomechanical properties as a prime example. Here we present a method based on photothermal off-resonance tapping (PORT) to increase the speed of such force spectroscopy measurements by at least an order of magnitude, thereby enabling high-throughput, quantitative nanomechanical mapping of a wide range of materials. Specifically, we use photothermal actuation to modulate the position of the AFM probe at frequencies that far exceed those possible with traditional actuation by piezo-driven *z* scanners. Understanding and accounting for the microscale thermal and mechanical behavior of the AFM probe, we determine the resulting probe position at sufficient accuracy to allow rapid and quantitative nanomechanical examination of polymeric and metallic materials.

**Introduction**

Atomic force microscopy (AFM)[1] is a valuable tool for acquiring nanomechanical maps of soft, solid and biological matter, such as sample elasticity, viscoelasticity, or adhesion and binding forces, along with topographic information at high spatial resolution under various environmental conditions[2–9]. Force-distance curve-based spectroscopy is the most commonly used measurement technique to acquire quantitative nanomechanical contrast, with material properties determined via the application of an appropriate contact mechanics model[2,10]. In traditional force spectroscopy, the tip-sample distance is modulated by a triangular waveform, which has the advantage of constant loading and unloading rates, but which introduces higher harmonics that can excite the AFM *xyz* scanner and system resonances. With typical *z* scanner resonances of several kHz, this comes at the cost of limiting the modulation rate at which force spectroscopy can be performed, to at best $\approx$ 100 Hz or $\approx$ 10 ms per force curve[2]. Performing an *xy* position grid of force spectroscopy curves to map the nanomechanical properties of a surface may thus take several minutes to an hour, depending on the resolution required. Although the highest reported force curve acquisition rates with AFM are up to several hundreds of kilohertz, these require a dedicated *z* scanner design optimized for speed, with severe constraints on the mechanical design and the range of the scanner[11,12]. To nonetheless achieve higher throughput, force spectroscopy data may be acquired using sinusoidal tip-sample distance modulation, thereby avoiding the excitation of higher harmonic system resonances and also more gently indenting the sample, as the tip velocity is reduced at turn-around points[13,14].

Nanomechanical mapping may also be performed via parametric methods such as phase imaging[15], contact resonance[16,17], and bimodal AFM[18–23], actuating AFM cantilevers at or close to their respective resonances. Although these methods may enable rapid nanomechanical mapping, the loading rates are restricted by the resonance frequencies that are being excited. The resulting resonant behavior invalidates the commonly assumed quasi-static relation between cantilever bending and sample elasticity, and requires a more complex modelling of tip-sample interaction forces[24,25].

These various limitations may be overcome by off-resonance force spectroscopy in which the tip-sample distance is modulated by direct actuation of the cantilever (instead of the *z* scanner)[26]. In this case, the maximum frequency for off-resonance excitation is limited by the cantilever's resonance, which is determined by its spring constant, inertia and damping[27,28]. The last two properties can be significantly affected by the surrounding medium, particularly in aqueous environments. Nevertheless, the resonance frequency of the AFM cantilever is typically much higher than that of the *z* scanner, even for high-speed AFM systems[29]. Direct cantilever actuation may be achieved, e.g., by piezo-acoustic[30,31], magnetic[32–34], resistive thermal[35,36] or electrostatic excitation[37,38].

Here, we favor photothermal excitation[39–41], directly actuating the cantilever by heating it with a laser. In brief, photothermal off-resonance excitation is based on an asymmetry in the cantilever heating and/or cantilever composition (e.g., metal coating on top of a silicon or silicon-nitride cantilever), causing the cantilever to act as a bimorph and to bend in response to such heating. Photothermal excitation stands out among other cantilever excitation methods[30–38] for being compatible with a wide range of commercially available probes and for being applicable in air and liquid environments[42].

Off-resonance photothermal actuation has previously been introduced for AFM imaging, also known as photothermal off-resonance tapping (PORT)[26], or as WaveMode in its commercial implementation. It has successfully been applied for rapid and gentle topography imaging of delicate samples such as proteins[26,43], living cells[44], virus capsids[45] and DNA three-point-star motifs[46,47].

In this study, we extend PORT from imaging to the mapping of nanomechanical properties. In doing so, we reveal non-trivial thermomechanical responses of the cantilevers and develop a calibration method to nonetheless enable the straightforward and high-throughput acquisition of quantitative nanomechanical data by photothermal excitation of AFM cantilevers.

**Results and Discussion**

**Photothermal actuation facilitates force curve acquisition over a wide range of frequencies**

As common in PORT, a first laser or superluminescent diode is used as a light source for detecting the cantilever by optical beam deflection onto a position-sensitive detector[48], and a second laser, typically focused close to the base of the cantilever, is modulated in intensity, thus providing a modulated heat source $Q$ for direct cantilever actuation [39–41,49] (Figure 1a). This facilitates the actuation of the cantilever with a smooth response over a frequency range from DC up to well beyond the cantilever resonance, here apparent as a sharp peak between 200 and 300 kHz (Figure 1b). For comparison, traditional force spectroscopy typically operates for ramp frequencies of up to ≈ 100 Hz, and piezo-actuated off-resonance tapping is limited by the resonance frequency of the $z$ scanner, typically not exceeding a few kHz for scanners with an operating range of a few microns or more.

In the absence of tip-sample forces, the cantilever deflection follows a sinusoidal movement when excited with a sinusoidal laser light modulation. When the probe tip intermittently contacts the sample surface, the cantilever deflects due to tip-sample forces. By subtracting the recorded free cantilever deflection from the recorded deflection with intermittent surface contact, the tip-sample interaction force signal can be reconstructed (Figure 1c). Using the cantilever deflection without surface contact as vertical probe position and the interaction signal as force, the signals can be transformed into a force versus position curve, where the absolute slope in the contact regime (here denoted as 'effective stiffness') provides a simple measure of the tip-sample stiffness (Figure 1d). More accurate estimates can be obtained by fitting contact mechanic models to the force-distance curve$^{2,4,5,10,50}$, as will be shown later.

The potential of WaveMode-based mechanical characterization can be exemplified by mapping the effective stiffness of a styrene-butadiene-styrene-polystyrene polymer blend (SBS-PS). As illustrated by a comparison of WaveMode and traditional spectroscopy operated at 25 kHz and 100 Hz excitation rate respectively, the vastly increased force-curve acquisition speed can be used, for example, to enhance the spatial pixel resolution of the nanomechanical maps without the need to increase the overall measurement time, while still yielding similar stiffness contrast (Figure 1e).

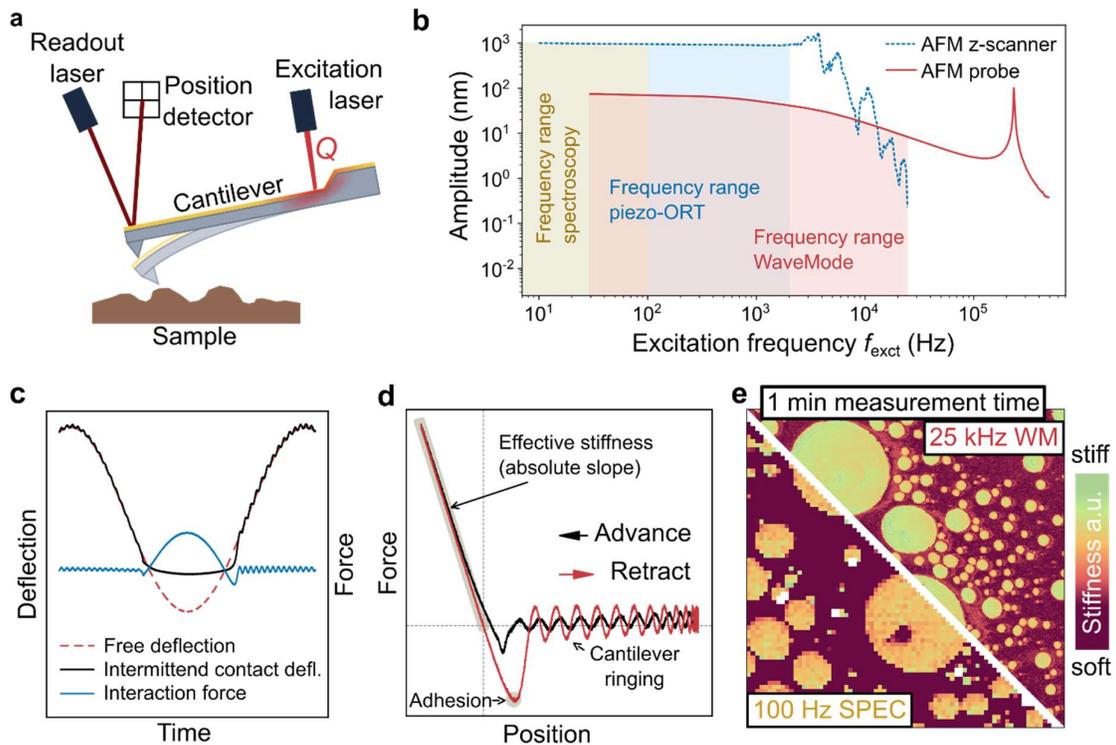

**Figure 1. Extending WaveMode imaging to nanomechanical mapping. (a)** Schematic of the applied WaveMode. The cantilever deflection is measured via the reflection of a readout laser onto a position-sensitive detector. A second, intensity-modulated laser locally heats and thereby actuates the cantilever, here at frequencies well below its mechanical resonance frequency. **(b)** Amplitude response of the AFM $z$ scanner and of the cantilever, as a function of frequency. Yellow, red and blue bands indicate the accessible frequencies at which force spectra can be recorded. **(c)** Cantilever deflection as a function of time for WaveMode with (intermittent contact deflection) and without (free deflection) intermittent surface contact. The tip-sample interaction force follows, after multiplication with the spring constant, from the difference between these two signals. **(d)** Force versus $z$ position reconstructed by plotting the interaction signal as a function of the free deflection. **(e)** Effective stiffness map (10 x 10 µm², arbitrary units) of a polymer blend (SBS-PS), demonstrating enhanced pixel resolution of WaveMode (WM, at 25 kHz pixel rate) compared with conventional spectroscopic mapping (SPEC, at 100 Hz) within the same measurement time.

**Understanding and correcting for hysteretic and scaling effects in WaveMode force curves**

For a quantitative assessment of the WaveMode-based force spectra, we first evaluated the spectra in air environment for the case where the stiffness of the sample (here: sapphire) vastly exceeded the stiffness of the AFM cantilever (Figure 2), and where viscous sample effects, hydrodynamic drag and plastic deformation could be neglected. In the contact region of the force curves, the cantilever deflection must then balance the attempted change in $z$ position of the cantilever, resulting in a slope of -1 nm/nm in the predicted deflection-versus-position curves. This would correspond to an infinite slope in a plot of the deflection against tip-sample separation.

For low excitation frequencies (0.1% of the cantilever resonance frequency; Figure 2, top) with the excitation laser positioned close to the base of the cantilever, the measured data follow the prediction of non-hysteretic behavior reasonably well, but the measured slope noticeably deviated from the predicted -1 nm/nm, deviating even more for excitation positions closer to the cantilever free end. At higher excitation frequencies (1% and 10% of the resonance frequency; Figure 2, center and bottom), such deviations from the expected slope were common for various excitation positions along the cantilever, and the measured curves showed non-trivial hysteretic behavior. Most remarkably, the advance curves (i.e., tip advancing to the sample) could lead or lag the retract curves depending on the excitation position, which is not expected for a purely elastic material. This hysteretic behavior and scaling effects in the force curves were qualitatively reproduced in force curves obtained from thermo-mechanical finite element method (FEM) simulations of a cantilever beam (Supplementary Figure 2), indicating the origin of the effect in the thermo-mechanical coupling. Finally, the experimental curves show oscillatory behavior in the non-contact region, as the cantilever resonance is excited upon the sudden release of the tip from the sample surface as the tip is withdrawn. This cantilever snap-out ringing effect has previously been reported for conventional piezo-driven off-resonance tapping[14,51] and is not specific to WaveMode. The deviations from the expected slope of -1 nm/nm indicate complications for extracting accurate mechanical data from the raw WaveMode curves. We conclude that the scaling inaccuracies and hysteretic behavior need first to be understood and next to be corrected by an appropriate calibration procedure.

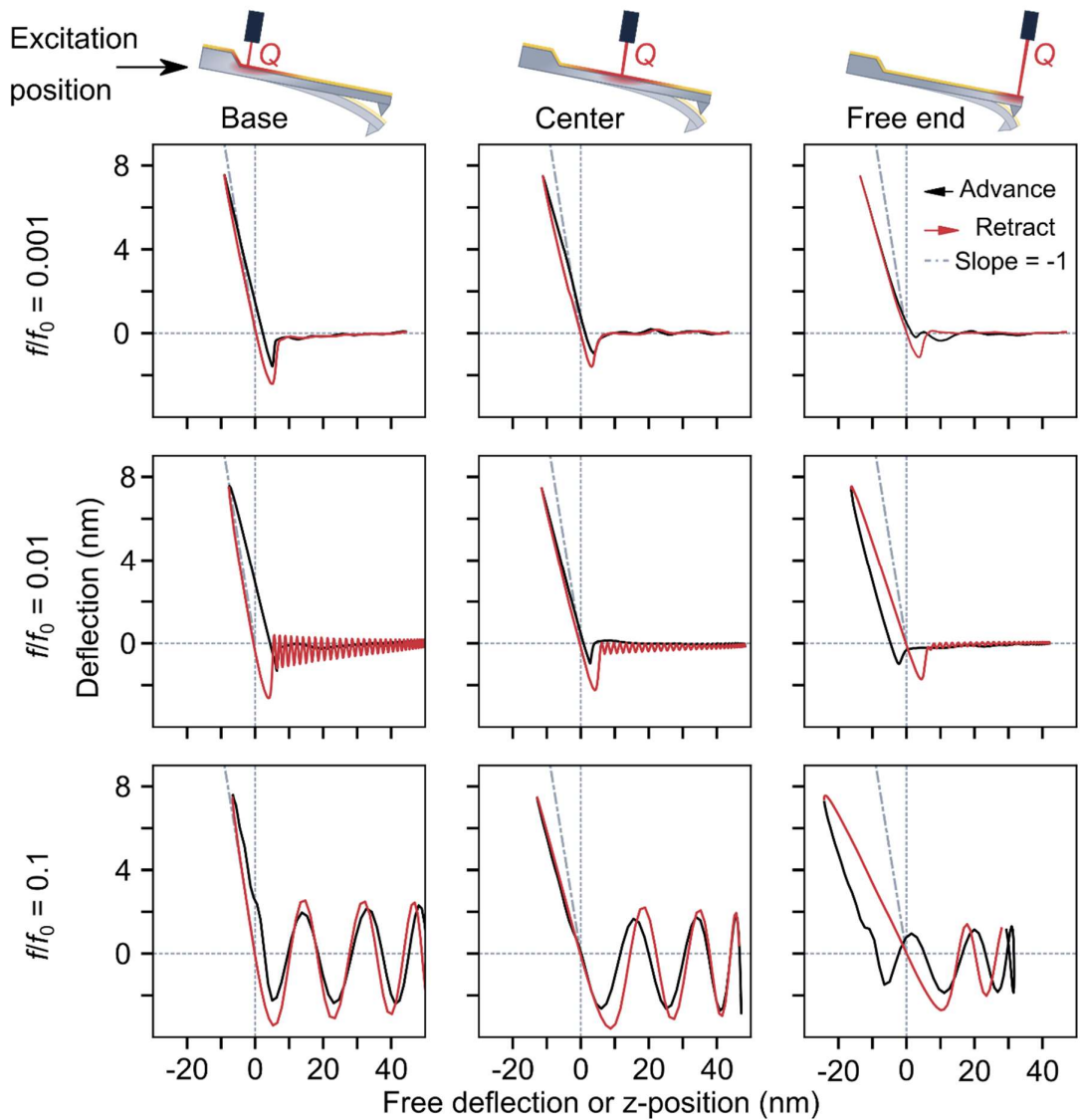

**Figure 2. Raw WaveMode force curves show strong dependencies on actuation position and frequency.** Raw deflection-versus-position curves obtained in WaveMode on a hard, sapphire surface, for (from left to right) excitation laser spot positions at the base, center and tip end, and for (from top to bottom) low to high excitation frequencies measured relative to the resonance frequency $f_0 = 249$ kHz of a probe with a spring constant $k_{probe} = 0.8$ N/m ± 0.16 N/m (WM0.8PTD, Nanosurf). Each panel shows the average of 10,000 cycles. The grey, dashed line shows the expected curve for the contact region ($z < 0$) of the data, with slope -1 nm/nm.

As described above, the cantilever serves as a sensor for measuring both tip-sample forces and tip-sample position in WaveMode. Cantilever deflections are read out using optical beam deflection, where a laser beam reflected from the cantilever surface lands on a position-sensitive detector (Figure 1a) which, together with the readout electronics, yields an electrical voltage proportional to the position of the laser spot on the detector $(V)$[48]. This voltage is usually translated into a cantilever displacement $(z)$ by scaling with the static deflection sensitivity $\sigma_s$, in nm/V,

$$z = V\sigma_s. \quad (1)$$

Of note, the reflection of the readout laser from the cantilever depends on the angle of the cantilever where the laser reflects from it, typically at the free end of the cantilever[52,53]. The translation of cantilever angle into cantilever position depends on the bending shape[50,54,55]. This is well known, and underpins the broadly applied correction factors for cantilever calibrations (e.g., $\sigma_{f0}/\sigma_s = 1.09$ for the first flexural mode of a rectangular cantilever), which account for differences in cantilever bending shape due to quasi-static tip-sample forces and bending shape due to resonant oscillation in the first flexural mode of the cantilever[53]. Other variations in deflection sensitivity might be expected when the free end of the cantilever becomes a fixed end by pushing/pinning it to a hard surface, but for practical tip-sample forces, these variations may usually be neglected[56].

These considerations prompted us to investigate the cantilever bending shape due to a thermal load as results from photothermal actuation. Using FEM simulations, we found that both the slope and the phase of the free end of the cantilever bending shape change with excitation position and frequency (Figure 3a). For a better visualization of the slope differences, the first derivatives of the bending shapes are plotted in Figure 3b. The bending shapes for off-resonance excitation differ from those observed when a point force is applied at the free end. This point force bending shape is typically assumed when calibrating the static deflection sensitivity $\sigma_s$ by measuring against a hard surface. Consequently, the deflection sensitivity due to photothermal excitation $\sigma$ will depend on excitation position and frequency, too. That is, for the position measurement on a photothermally actuated cantilever, we need to introduce a new deflection sensitivity

$$\sigma(x_{\text{exc}}, f_{\text{exc}}) = a_{\text{corr}}(x_{\text{exc}}, f_{\text{exc}})\, e^{i\varphi_{\text{corr}}(x_{\text{exc}}, f_{\text{exc}})}\, \sigma_s, \quad (2)$$

which differs from the static deflection sensitivity $\sigma_s$ by a scaling factor $a_{\text{corr}}(x_{\text{exc}}, f_{\text{exc}})$ and a phase shift $\varphi_{\text{corr}}(x_{\text{exc}}, f_{\text{exc}})$. The relation between a given cantilever displacement and the angular beam deflection at the readout position depends on mechanical response and on the thermal dynamics of the heat wave traveling through the cantilever[57]. The deflection sensitivity scaling is necessary since the cantilever bending shape and therefore the angular deflection at the cantilever's free end generally depends on the excitation frequency $f_{exc}$ (Figure 1b) and excitation position $x_{exc}$[49,57,58]. The phase correction becomes necessary, since there is a phase difference between the cantilever displacement and angular deflection, caused by the thermal wave having its own dynamics propagating through the beam, which defines the elastic beam response[59]. Hence, as determined from the free deflection of a photothermally excited cantilever oscillating above a sample surface, the corrected tip position is

$$z_{corr} = V_{free}\sigma(x_{exc}, f_{exc}). \quad (3)$$

By contrast, the interaction signal is the difference between measured deflection in intermittent contact and measured deflection of the free wave. Since this difference is entirely due to the tip-sample interaction, the according change in cantilever position follows the prediction for a point-source force acting at the free end of the cantilever (Figure 3a,b), which is independent of frequency if cantilever resonances are avoided. This follows from Euler-Bernoulli theory, in which the overall cantilever response due to different forces is the linear superposition of the respective cantilever responses to these forces[60]. Hence, disregarding the more complex cantilever response due to photothermal excitation, the tip-sample force $F_{ts}$ can simply be calculated from the difference between free and intermittent-contact deflections (in Volts), first multiplied by the static deflection sensitivity for conversion into a change in cantilever position, and next multiplied by the cantilever spring constant $k_{probe}$ for conversion into Newtons. That is,

$$F_{ts} = (V_{interm} - V_{free})\,\sigma_s\,k_{probe}. \quad (4)$$

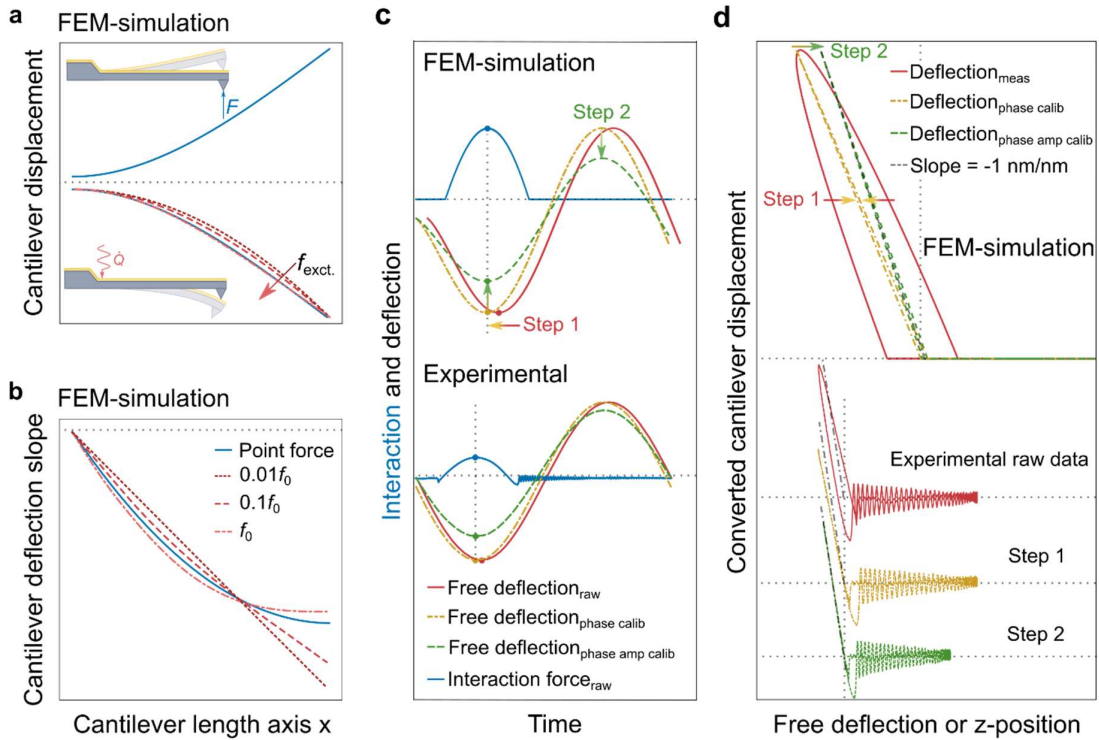

**Figure 3. Calibrating WaveMode spectroscopic data. (a)** FEM simulations of cantilever bending shapes deformed by a point force at the free end and a heat source at the base, representing tip-sample forces and photothermal excitation respectively. The bending shape caused by thermal strain changes with excitation frequency. **(b)** The first derivatives of the bending shapes shown in (a). The output of an optical beam deflection measurement scales with this slope close to the end of the cantilever. **(c)** Step 1 of the calibration: aligning the minimum of the free deflection and the maximum of the interaction peak on a hard surface to remove phase shift. **(d)** Step 2 of the calibration: scaling the free deflection such that the slope of the in-contact part of the deflection-position curve equals -1 nm/nm, as expected on a hard surface. The determined phase/temporal shift and amplitude scaling are then applied to all free deflection data as recorded in a given experimental setup.

The challenge of obtaining accurate force curves by WaveMode is therefore reduced to finding a scaling factor $a_{\text{corr}}(x_{\text{exc}}, f_{\text{exc}})$ and a phase shift $\varphi_{\text{corr}}(x_{\text{exc}}, f_{\text{exc}})$, and applying these to determine a calibrated free deflection signal (Figure 3c). These numbers can be obtained via various routes, one of which is by measuring WaveMode-based force curves against a hard (e.g., sapphire) surface, first adjusting $\varphi_{\text{corr}}(x_{\text{exc}}, f_{\text{exc}})$ to minimize hysteresis, effectively aligning the minimum of the free deflection signal to the maximum of the interaction signal, and next adjusting $a_{corr}(x_{exc}, f_{exc})$ to obtain the predicted slope of -1 nm/nm (Figure 3c,d). This procedure should allow for calibrated WaveMode forces within a given experimental setup including a given photothermal excitation position $x_{exc}$. Where needed, such calibration measurements can be done over a range of frequencies $f_{exc}$. A comprehensive overview of deflection sensitivity phase and scaling correction factors is shown as a bar plot in Supplementary Figure 3 and as table in Supplementary Table 1. An animated overview of the cantilever bending behavior is provided in Supplementary Video 1.

**Calibrated WaveMode force spectroscopy yields accurate estimates of sample stiffness**

For experimental validation of this calibration method, we selected a second tipless cantilever as a reference sample, because it has a predictable variation in its stiffness depending on whether it is probed closer to its support chip or closer to its free end (Figure 4a,b). For frequencies that are small compared to the cantilever resonance frequencies involved, such a reference sample is expected to yield a linear force curve in the contact regime, with an absolute slope $s(x_{ref})$ of

$$s(x_{ref}) = \frac{k_{ref}(x_{ref})}{k_{probe}+k_{ref}(x_{ref})} \quad , \quad (5)$$

where $k_{probe} = 0.68$ N/m $\pm$ 0.14 N/m is the stiffness of the cantilever used to measure the force curve (WM0.6AuD, Nanosurf, $f_{0,probe} = 222$ kHz) and $k_{ref}(x_{ref})$ is the stiffness of the cantilever acting as a reference sample (USC-f1.5-k0.6, Nanoworld, $f_{0,ref} = 2.081$ MHz). The reference stiffness $k_{ref}(x_{ref})$ is expected to scale with the measurement position $x_{ref}$ along the reference cantilever as $x_{ref}^{-3}$ (Ref. [61]), diverging as $k_{ref}(x_{ref} = 0) \to \infty$ and approaching 0.62 N/m $\pm$ 0.12 N/m as measured using the Sader method[62] at the free end $x_{ref} = L_{ref}$. Accordingly, the absolute slope $s(x_{ref})$ is expected to range from 1.0 to 0.5 $\pm$ 0.1 for measurements taken at different positions between the base and the free end of the reference cantilever. Using the measurement close to $x_{ref} = 0$ for the hard-sample calibration as described in the previous section we find that the calibrated WaveMode curves are matching the predicted trend of a declining absolute slope already when probing towards the free end of the reference cantilever (Figure 4c).

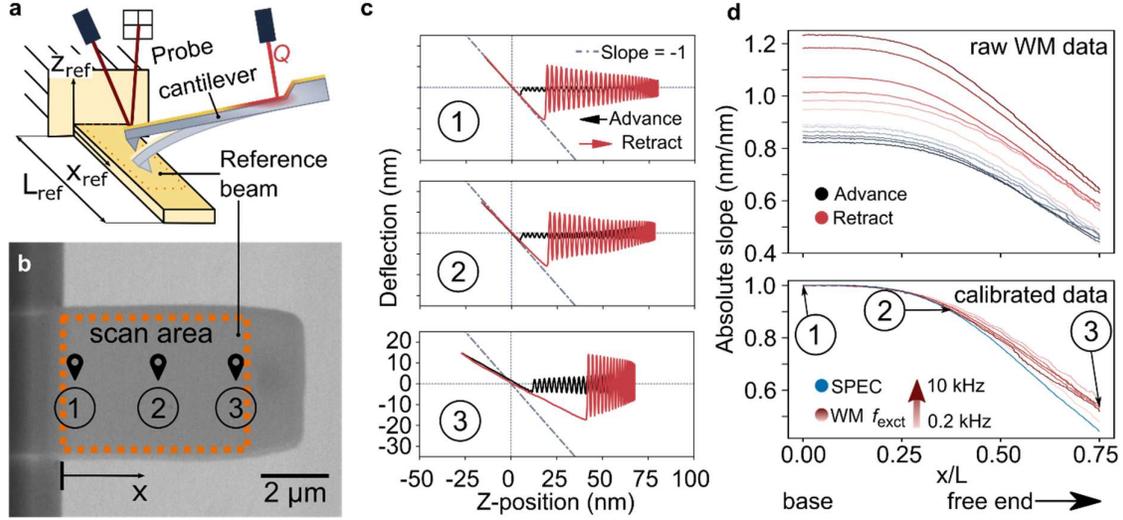

**Figure 4. Validation of the calibration procedure. (a)** Schematic (not to scale) of a validation measurement using WaveMode with a probe cantilever to detect local stiffness of a reference cantilever beam that acts as a sample. **(b)** SEM image of the reference beam (USC-F1.5-k0.6, $f_0$ = 2.081 MHz, $k_{ref}$ = 0.62 N/m ± 0.12 N/m Nanoworld AG), highlighting 3 different x-positions at which conventional spectroscopy (SPEC, 50 Hz z scanner actuation) and WaveMode (WM, at 0.2, 0.5, 1, 2, 5 and 10 kHz) data were recorded. The stiffness and the measured absolute slope declines towards the free end of the reference beam, as the beam is more compliant at larger distances from its fixed end. **(c)** Calibrated deflection-position curves from the 2 kHz WaveMode measurement selected from base, center and free-end position of the reference beam, showing the change in slope. **(d)** Measured slope of the in-contact part of the measured force curves along the normalized length axis x of the reference beam for different excitation frequencies using WaveMode (WM) and conventional spectroscopy (SPEC). The expected absolute slope is 1.0 nm/nm at the base (hard surface) and around 0.5 ± 0.1 nm/nm at the free end of the reference cantilever, as determined from the spring constants of the respective cantilevers.

For a more quantitative analysis, we performed WaveMode measurements in the range of $0 < x_{ref} < 0.75 L_{ref}$, with excitation frequencies $f_{exct} = 0.2 – 10$ kHz, we compared raw and calibrated data (Figure 4d). Briefly, the raw data show ±20% scatter in the measured absolute slopes for different excitation frequencies and considering both advance and retract curves. By contrast, when $a_{corr}(x_{exc}, f_{exc})$ and $\varphi_{corr}(x_{exc}, f_{exc})$ were determined from the measurements at $x_{ref} = 0$, all data collapsed onto a single master curve, from 1.0 nm/nm for $x_{ref} \approx 0$ (not unexpected given the calibration procedure against a hard surface for every used excitation frequency) within 23% to the expected 0.7 nm/nm for $x_{\text{ref}} = 0.75\, L_{\text{ref}}$. Of note, the calibrated WaveMode results also match data acquired by traditional, linearly ramped force spectroscopy to within 22% or better, depending on the applied excitation frequency. In conclusion, when calibrated by the proposed procedure, WaveMode force curves allow to determine the mechanical properties of a model, reference sample to within a confidence interval similar to what generally applies to nanomechanical data, considering errors in deflection and spring constant calibration[63,64] as well as inaccuracies of indentation models used to extract mechanical properties[65,66].

**WaveMode enables high-throughput and high-resolution nanomechanical analysis of polymer and metal surfaces**

Having established the quantitative nature of WaveMode nanomechanical analysis on a model sample, we next sought to demonstrate its applicability to the high-throughput, high-resolution mechanical analysis of materials, with a polymer blend and a soft metal as examples.

The polymer blend consisted of two components, the rubber-like elastomer styrene-butadiene-styrene (SBS) and the hard and brittle polystyrene (PS), casted as a thick (> 2 µm) film on a glass slide. Acquiring WaveMode force spectra at 25 kHz, with a WM0.8PTD probe ($f_0$ = 253 kHz, $k_{\text{probe}}$ = 0.81 N/m ± 0.16 N/m), we obtain high pixel resolution (420 × 420 pixels) topographic scans of the SBS-PS surface in 1 min per frame (Figure 5a). While this is a relatively standard throughput for topographic imaging, the WaveMode data acquisition and calibration has the advantage of providing force curves at the same rate and spatial resolution: fitting these data with a DMT contact-mechanics model[67] yields an according elasticity (Young's modulus) map, and an adhesion map is readily determined based on the minimum force in each retract curve.

Displayed on a logarithmic scale to emphasize the wide spread in Young's moduli, the elasticity map shows a separation of three phases, with two softer phases at 46 ± 7 and 91 ± 39 MPa and a stiffer phase at 0.9 ± 0.2 GPa (mean ± standard deviation from fitting a gaussian distribution to the histogram peaks). In the adhesion map, the softer phases are also clearly distinct from the stiffer phase by higher adhesion forces ($F_{\text{ad}}$ = 12 nN versus 7 nN). We assign the soft and sticky phases to higher local SBS concentration, and the stiff and less sticky domains to higher local PS concentration. For reference, the bulk Young's Modulus of the polymer blend depends on the concentration of its components and is approximately at around 0.6 GPa for 50% SBS content[68]. A scanner-based spectroscopy measurement at a 1250 x lower indentation rate of the same sample area shows a stiffer phase at 1.5 ± 0.7 GPa (Supplementary Figure 4a). The two softer phases clearly distinguishable in the WaveMode measurement are barely visible in the spectroscopy map with the two phases overlapping in a single peak at 12 ± 4 MPa in the histogram.

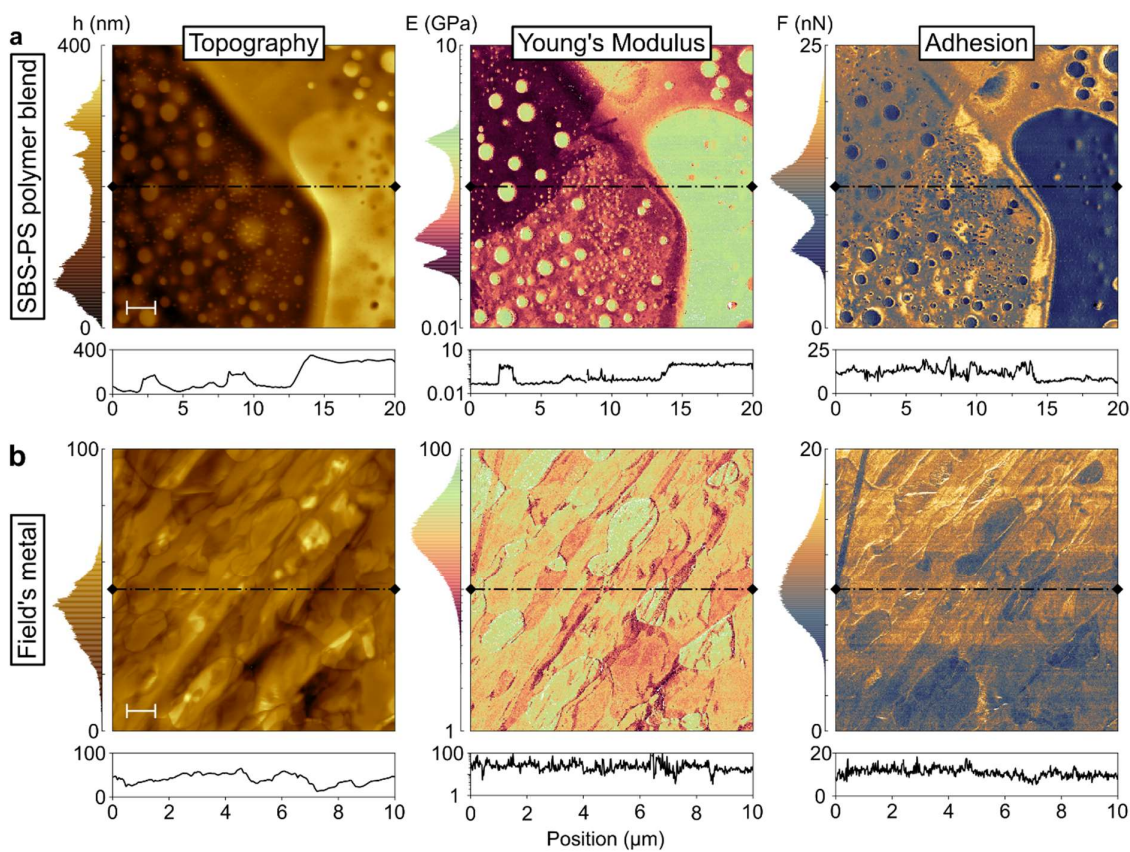

**Figure 5. WaveMode nanomechanical analysis on heterogeneous Samples.** Maps, histograms, and line profiles of **(a)** a polymer blend (SBS-PS), scale bar 2 µm, 420 x 420 pixels and **(b)** a soft metal alloy consisting of indium, bismuth and tin (Field's metal), scale bar 1 µm, 500 x 500 pixels. Both measurements were acquired at an excitation frequency of 25 kHz, with a scan time of (a) 1 min and (b) 1.2 min, with the WaveMode calibration done against a hard sapphire surface.

The soft metal sample is Field's alloy, an eutectic alloy consisting of 51% indium, 32.5% bismuth, indium, and 16.5% tin[69] with a melting temperature of 62 °C[70] and a bulk Young's Modulus of 9.25 GPa[71], an order of magnitude larger than the stiffest value measured on the SBS-PS polymer blend. To facilitate measurements in this higher stiffness range, a stiffer, WM20PTD probe was used ($f_0$ = 1.24 MHz, $k_{probe}$ = 32 N/m ± 6 N/m).

Also, on this sample (Figure 5b), three phases can be identified, although overlapping in the histograms, with a mean value of 28 ± 15 GPa. This is larger than the expected Young's modulus and larger than the mean value of 9 ± 3 GPa determined by traditional force spectroscopy data acquired at the same location, but at 1000× lower rate (Supplementary Figure 4b). Hence, we attribute the larger stiffness in the 25 kHz WaveMode measurement to load rate dependent effects, which was demonstrated for the bulk material of Field's metal exhibiting viscoelastic-viscoplastic behavior[72].

Both measurements demonstrate the capacity of WaveMode-based spectroscopy for high-throughput, high-resolution nanomechanical mapping of surfaces, not limited by the mechanical properties of the $z$ scanner. With a < 5 kHz $z$ scanner as is the case here and quite generally for standard z scanners in commercial AFM instruments, a 25 kHz WaveMode measurement results in a 10× to 25× enhancement in possible pixel rate and thereby possible speed for nanomechanical mapping.

**Discussion and Conclusion**

As AFM depends on collecting data by scanning a mechanical probe across a sample, it is subject to various mechanical factors limiting its throughput. Over the past decades, there have been many developments advancing the speed at which AFM data can be acquired, most notably in imaging by high-speed AFM[73,74]. These developments also benefit nanomechanical mapping and force spectroscopy, but $z$ scanner mechanics remains a crucial factor. Specifically, these mechanics impose the need for a trade-off between speed on one hand, which is more easily achieved with a small-range scanner, and $z$ range on the other, where a larger scanner range facilitates the applicability of AFM on a wider range of samples. Here, we have reported on a route to circumvent the limitations due to $z$ scanner mechanics (up to few kHz in many common commercial systems), demonstrating photothermally actuated force spectroscopy and nanomechanical mapping at frequencies up to 25 kHz (Figure 1,5). This increased measurement speed will enable the monitoring of dynamic processes, such as changes in nanomechanical sample properties, appearing at time scales of tens of seconds.

This effort builds on previous work on photothermal actuation in AFM[39–41], and particularly on PORT-based AFM imaging[26]. Of note, photothermal actuation has previously been proven beneficial for higher-speed single-molecule force clamp measurements[75], nanorheological measurements[76,77], shear stress studies of polymers[78] and biological cell mass measurements[79]. In our approach, applicable in gaseous as well as in liquid environment[26,43–47], the only remaining mechanical constraint is due to the microfabricated cantilever itself. Force curves can practically be acquired up to ≈10% of the cantilever resonance frequency in air, beyond which cantilever ringing (as visible in Figure 2) severely affects the quality of the spectra, and beyond which inertial effects may induce significant error[27]. With stiffer (e.g. the WM20PTD probe used in Figure 5b) and/or smaller[74] cantilevers with resonance frequencies ≥ 1 MHz, this enables force spectrum acquisition at ≥ 100 kHz, entering the realm of high-speed AFM.

At this point, the remaining speed limitations are electronic and computational, with ample scope for improvement. First, the measurement of cantilever deflection here occurred at a rate of 3 MHz, such that at a 100 kHz force curve would contain only 15 data points in one direction (advance or retract). Second, at such rates, data transfer and storage also become more challenging. In addition, when maximizing $xy$ scan rates accordingly, it will also require further optimization of $z$-feedback[73,80].

Besides demonstrating high-throughput nanomechanical mapping, we have revealed that the photothermal actuation leads to (at first sight) non-trivial force curve appearance, with scaling inaccuracies and hysteretic effects, both of which depending on excitation frequency and position of the photothermal actuation laser (Figure 2). These effects can be attributed to the thermomechanical response of the cantilever to the photothermal actuation. By recognizing that these only affect the $z$ position measurement in the force curves and not the force measurement itself, we have designed, implemented, and tested a calibration method (Figure 3) that enables photothermally actuated nanomechanical mapping that is not only fast, but also quantitatively accurate (Figure 4). We have used this ability to acquire nanomechanical mapping at high pixel resolution on polymer and metal surfaces (Figure 5). This method relies on the superposition of forces principle for Euler-Bernoulli beam theory, which is valid for small beam deflection angles and linear material behavior. Both assumptions are generally applicable to AFM measurements, with resultant uncertainties on par with other factors present in AFM material property measurements (e.g. tip geometry and spring constant). For more accurate quantification, models incorporating non-linear cantilever bending, material property effects, and consideration of further thermal boundary conditions such as heat transfer through the tip into the sample and convection along the probe may be needed.

In conclusion, we have presented a method for fast and quantitative acquisition of nanomechanical properties by off-resonance cantilever actuation via photothermal actuation, with up to ten times faster throughput compared to traditional AFM-based nanomechanics. The proposed calibration method, supported by findings from thermomechanical FEM-simulations and validated by a measurement of a reference beam, enables nanomechanical results that are not only fast, but also quantitatively correct. Altogether, this opens a path to further develop nanomechanical materials characterization to achieve orders-of-magnitude increase in throughput while retaining quantitative results.

# Methods

## AFM Instrumentation

All measurements were performed using a DriveAFM system with a CX controller (Nanosurf). The WaveMode NMA mode of the control software (Nanosurf Studio) was used to acquire the WaveMode cycles at a sampling rate of 3 MHz.

A wide range of commercially available probes featuring a reflective metal coating are compatible with photothermal excitation and WaveMode. However, the optimal performance can be achieved by using small cantilevers with a high resonance frequency to enable high speeds, and with a soft spring constant to allow the control of gentle contact forces, and a reflective metal coating optimized for efficient photothermal excitation to reach large amplitudes. WM0.6AuD probes (Nanosurf, $f_0$ = 0.3 MHz and $k_{probe}$ = 0.6 N/m) were used for reference cantilever probing. WM0.8PTD probes (Nanosurf, $f_0$ = 0.25 MHz and $k_{probe}$ = 0.8 N/m) were employed for polymer measurements, and WM20PTD probes (Nanosurf, $f_0$ = 1.2 MHz and $k_{probe}$ = 20 N/m) were utilized for Field's metal alloy measurements. For the cantilevers used in this work, photothermal actuation was implemented over a range of 1 kHz to 1 MHz, achieving amplitudes of 10~100 nm (Supplementary Figure 1).

## Data analysis and processing

AFM image processing was performed using Gwyddion (Version 2.62)[81], applying median of difference and plane fit for topography levelling. Phase and amplitude calibration factors were determined from measurements against a hard surface using a custom Python script, and data plotting was conducted using additional Python scripts. Deflection sensitivity (force curves measurements against hard surface), spring constant calibration (Sader method)[62], and force curve analysis were carried out using built-in tools in Nanosurf Studio. We applied a 20% measurement uncertainty to the measured spring constant, based on conservative estimates for dynamic experimental spring constant calibration methods[63]. The Young's Moduli of the polymer and metal sample were determined from fitting a so-called DMT-model[67] into the retract force-distance curves using the following fit function, with $E_{eff}$ being the effective Young's Modulus, $R_t$ the tip radius, $d_{ts}$ the tip-sample distance and $\gamma$ the surface energy:

$$F_{ts} = \frac{4}{3}E_{eff}\sqrt{R_t d_{ts}^3} - 4\pi R_t \gamma \qquad (6)$$

From the effective Young's Modulus $E_{eff}$, the sample Young's Modulus $E_s$ can be calculated using the Young's Modulus of the tip $E_t$ and the Poisson ratios of tip and sample $\nu_t$, $\nu_s$:

$$\frac{1}{E_{\text{eff}}} = \frac{1-\nu_t^2}{E_t} + \frac{1-\nu_s^2}{E_s} \tag{7}$$

Nominal values from the cantilever manufacturer were taken for the tip geometry (10 nm tip radius) and a value of 280 GPa for the Young's modulus and 0.25 for the Poisson ratio of the silicon tip was used. A Poisson ratio of 0.35 was chosen for the analysis of the data obtained from the SBS-PS polymer blend and Field's metal measurement.

**Computational modeling**

A 2D finite element method (FEM) beam model was developed in COMSOL Multiphysics (Version 6.3), incorporating both thermal and mechanical modeling. The model was used to simulate the cantilever bending shapes under various excitation conditions, including harmonic temperature disturbance to simulate photothermal excitation and a point force at the tip end to simulate tip-sample interactions. An overview of the model parameter used is shown in Supplementary Table 2. The simulated cantilever consists of a silicon nitride beam ($k = 1.16$ N/m, $f_0 = 259.5$ kHz) with a thinner gold layer on top and is attached to a bulky silicon nitride unit to take heat transfer into the probe chip into account. The top metal layer is divided into ten equally spaced units, where always three units were selected to simulate the position of heat introduced at base, center and free end of the beam.

Mechanical loads, and initial and boundary condition: the system consists of a one side clamped beam, attached to the bulk unit representing the probe chip which is fixed. The beam is initially undeformed. A point force following the characteristic curve of a mechanical spring at the free end of the beam represents tip-surface interaction forces and is turned on and off with an event to simulate the intermittent contact of the tip.

Thermal loads, and initial and boundary conditions: the beam and its environment were at an initial temperature of 20°C. The chip and the beam were thermally isolated against the surrounding medium (air) except for a region where a sinusoidally varied temperature disturbance was introduced to simulate photothermal excitation. No convection along the beam was assumed.

The simulated cantilever bending shapes of Figure 3a,b were retrieved from a frequency domain simulation using the maximum amplitude for each frequency, not considering phase shift here. The bending signals in Figure 3c,d were obtained from a time dependent simulation of a cycle of harmonic temperature disturbance at the base of the beam.

**Sample preparation**

As reference cantilever in Figure 4, an USC-F1.5-k0.6 cantilever (Nanoworld AG) was used and glued onto a silicon substrate using two-component epoxy (Araldite). The tip of the reference cantilever was previously removed using a focused ion beam. The nominal specifications of this cantilever given by the manufacturer are $k_{\text{ref}} = 0.6$ N/m, $f_0 = 1.5$ MHz, 7 x 3 x 0.1 µm$^3$ (length x width x thickness).

Styrene-Butadiene-Styrene (SBS) and Polystyrene (PS) granulate particles were dissolved in toluene at a concentration of 10 mg/ml. The solutions were mixed in a 50:50 ratio and dispensed onto a glass slide using a glass pipette. The polymer blend film formed on the glass slide within a few minutes.

A piece of approximately 5 g from a Field's metal ingot (Eugen Müller) was melted at $T = 70$ °C ($T_s = 62$ °C) and deposited onto a glass slide placed on a hot plate. The molten metal was then squeezed with a second glass slide. Both glass slides were immersed in water to accelerate the cooling process. After solidification, the glass slides were separated, leaving a flat piece of Field's metal with a smooth surface on one of the glass slides.

**Acknowledgements**

It is a pleasure to thank Dominik Ziegler (Nanosurf), Cornelia Pichler (KIT), and Richard Thelen (KIT) for support and useful discussions. We would like to thank Marcus Wyss from the Nano Imaging Lab, Swiss Nanoscience Institute (SNI), for his support with FIB operation and SEM imaging and acknowledge Justine Nyarige (KIT) and Alban Muslija (KIT) for EDX measurement of the Field's metal. Furthermore, we acknowledge support of the Karlsruhe Nano Micro Facility (KNMFi, www.knmf.kit.edu), a Helmholtz Research Infrastructure at Karlsruhe Institute of Technology (KIT, www.kit.edu).

# Supporting Information

**High-speed quantitative nanomechanical mapping by photothermal off-resonance atomic force microscopy**

*Hans Gunstheimer[1,2], Gotthold Fläschner[2,†], Jonathan D. Adams[2], Hendrik Hölscher[1,3, *], Bart W. Hoogenboom[2, *]*

[1] Institute of Microstructure Technology, Karlsruhe Institute of Technology (KIT), Hermann-von-Helmholtz-Platz 1, 76344 Eggenstein-Leopoldshafen, Germany  [2] Nanosurf AG, Gräubernstrasse 12 – 14, 4410 Liestal, Switzerland

[3] Karlsruhe Nano Micro Facility (KNMFi), Karlsruhe Institute of Technology (KIT), Hermann-von-Helmholtz-Platz 1, 76344 Eggenstein-Leopoldshafen, Germany

[†] Present address: Institute for Bioengineering of Catalonia (IBEC), C/ Baldiri Reixac 10-12 · 08028 Barcelona, Spain.

* Corresponding authors. Email: hendrik.hoelscher@kit.edu, hoogenboom@nanosurf.com

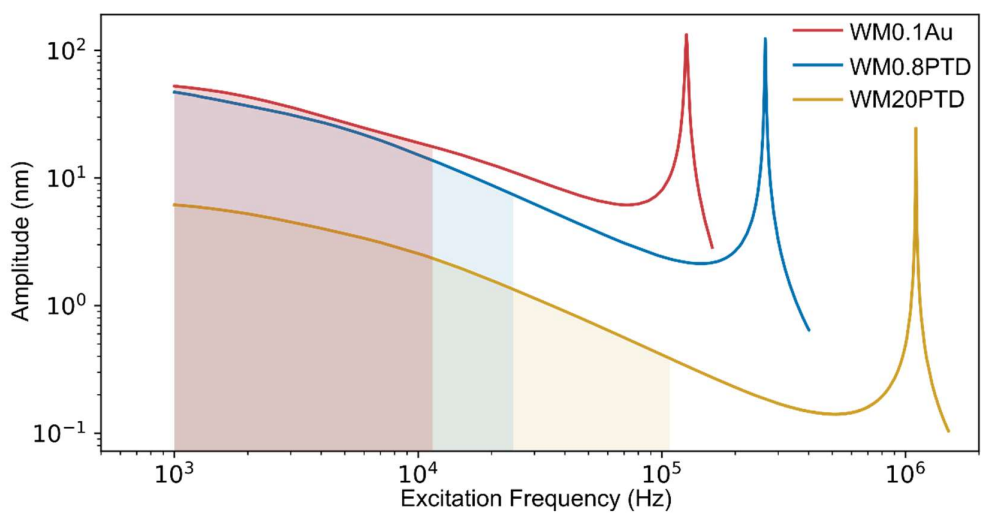

**Supplementary Figure 1. Frequency response of different AFM probes (WM0.1Au, WM0.8PD and WM20PTD) to photothermal excitation at 1 mW DC and 0.5 mW AC power.** The amplitude excitation efficiency of a cantilever can be enhanced either by lowering the cantilever stiffness (for similar cantilever width) or optimizing the reflective coating. WM0.8PTD is softer than WM20PTD and reaches higher amplitudes for the same laser power, while the WM0.8PTD is stiffer than the WM0.1Au probe but reaches similar amplitudes thanks to optimized cantilever coating. The filled areas show the suggested frequency ranges (up to 10% of the probe resonance frequency $f_0$) for WaveMode operation in air.

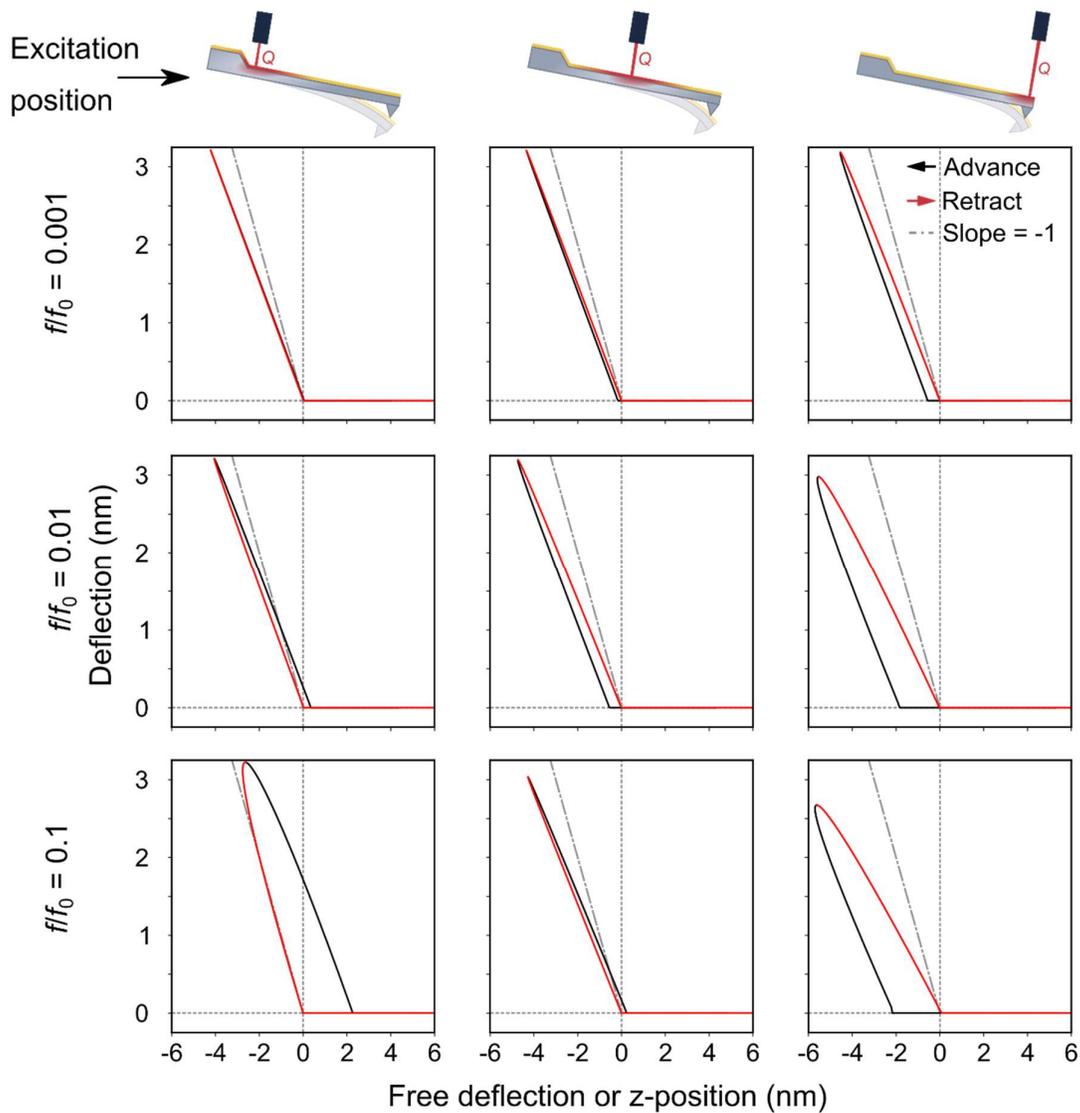

**Supplementary Figure 2. Raw deflection-versus-position curves obtained from cantilever beam FEM-simulations with a sinusoidally disturbed temperature at the base, center and free end of the beam.** A point force, with its magnitude increasing with deformation at the free end and having a force constant 1000 times higher than the beam spring constant to simulate a hard surface, was used to simulate tip-sample interaction forces. The grey, dashed line shows the expected curve for the contact region ($z < 0$) of the data, with an absolute slope of 1 nm/nm.

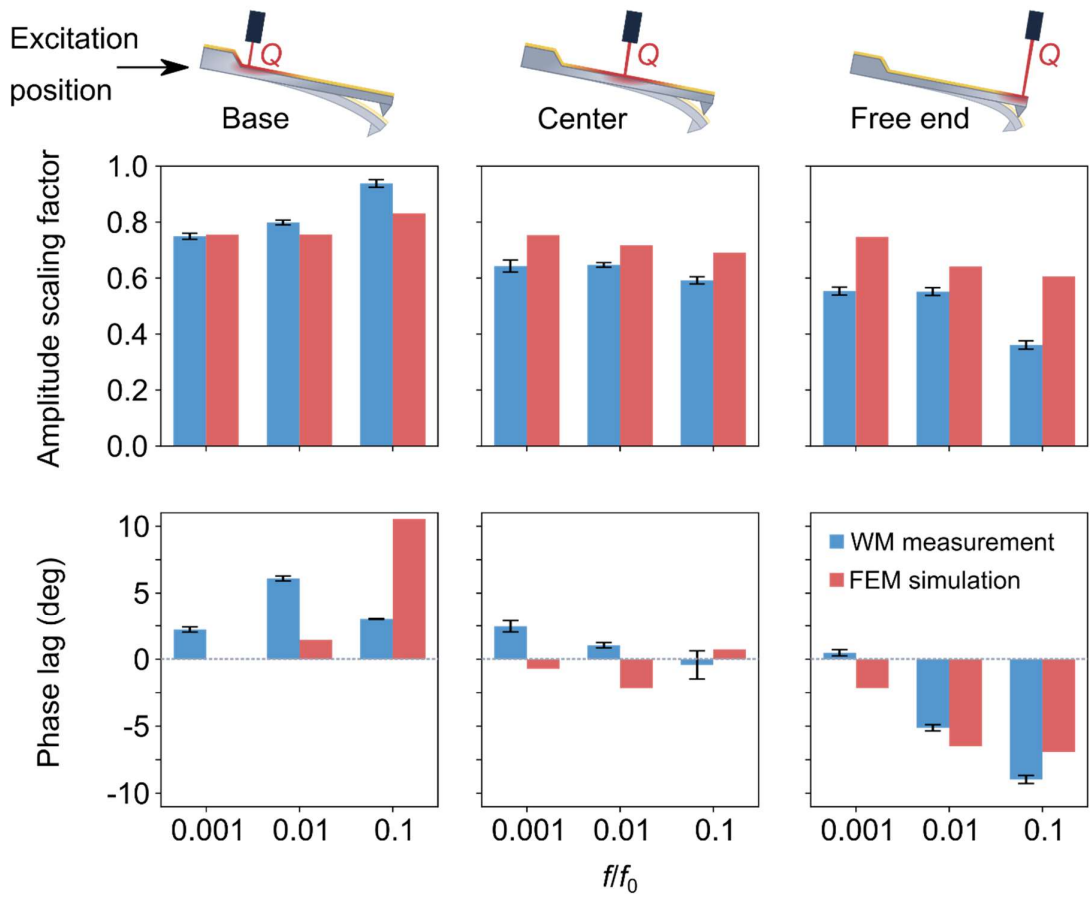

**Supplementary Figure 3.** Amplitude and phase correction factors obtained from the measurement data against a hard surface shown in Fig. 3 and cantilever beam FEM-simulation.

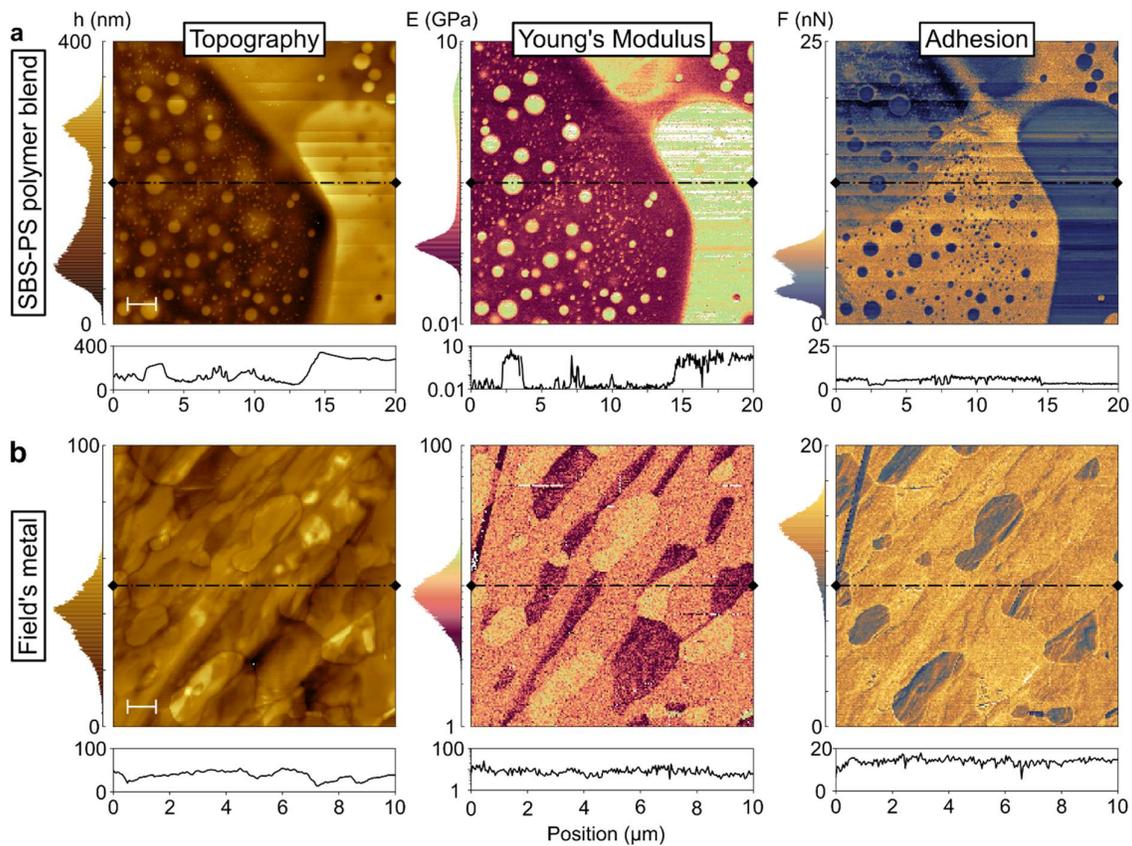

**Supplementary Figure 4. Spectroscopy nanomechanical analysis on heterogeneous Samples.** Maps, histograms, and line profiles of **(a)** a polymer blend (SBS-PS), scale bar 2 μm, 420 x 420 pixels and **(b)** a soft metal alloy consisting of indium, bismuth and tin (Field's metal), scale bar 1 μm, 200 x 200 pixels. Both measurements were acquired at an excitation frequency of 20 Hz (SBS-PS measurement) and 25 Hz (Field's metal measurement), with a measurement time of (a) 147 min and (b) 27 min.

**Supplementary Table 1.** Amplitude and phase correction factors obtained from the measurement data against a hard surface shown in Fig. 3 and cantilever beam FEM-simulation.

| Excitation position | Excitation frequency | Phase lag meas. (deg) | Phase lag sim. (deg) | Amplitude scaling factor meas. | Amplitude scaling factor sim. |
|---|---|---|---|---|---|
| Base | 0.1% $f_0$ | 2.2124 | 0 | 0.7495 | 0.7550 |
| | 1% $f_0$ | 6.1002 | 1.443 | 0.7983 | 0.7558 |
| | 10% $f_0$ | 2.9970 | 10.558 | 0.9382 | 0.8311 |
| Center | 0.1% $f_0$ | 2.4554 | -0.720 | 0.6430 | 0.7536 |
| | 1 % $f_0$ | 1.0404 | -2.164 | 0.6471 | 0.7175 |
| | 10% $f_0$ | -0.4230 | 0.729 | 0.5919 | 0.6907 |
| Free end | 0.1% $f_0$ | 0.4780 | -2.160 | 0.5537 | 0.7473 |
| | 1% $f_0$ | -5.1228 | -6.492 | 0.5517 | 0.6411 |
| | 10% $f_0$ | -8.9730 | -6.924 | 0.3608 | 0.6057 |

**Supplementary Table 2.** Parameter of the thermo-mechanical cantilever beam FEM simulation.

| Property | Chip | Beam | Metal Layer |
|---|---|---|---|
| Size (μm) | 100 x 10 | 50 x 0.5 | 50 x 0.15 |
| Mesh | Free Quad | Mapped | Mapped |
| Maximum element size (μm) | 10 | 0.5 | 0.5 |
| Minimum element size (μm) | - | 0.045 | 0.045 |
| Material | Silicon Nitride | Silicon Nitride | Gold |
| Density (kg/m3) | 3100 | 3100 | 19300 |
| Young's modulus (GPa) | 250 | 250 | 70 |
| Poisson's ratio | 0.23 | 0.23 | 0.44 |
| Thermal expansion coefficient (1/K) | 2.3E-6 | 2.3E-6 | 14.2E-6 |
| Heat capacity (J/(kg·K)) | 700 | 700 | 129 |
| Thermal conductivity (W/(ms$^3$K)) | 20 | 20 | 317 |

**Supplementary Video 1.**